\documentclass[11pt, letterpaper]{article}

\usepackage{amssymb, amsthm, amsmath, bm, color}
\usepackage{natbib}
\usepackage{graphicx}
\usepackage[figuresright]{rotating}
\usepackage{subfig}
\usepackage{soul}
\usepackage{multirow}
\usepackage{fullpage}
\usepackage{setspace}
\usepackage{pdflscape}
\usepackage[table]{xcolor}
\usepackage[hidelinks]{hyperref}
\usepackage{longtable,booktabs,threeparttablex}
\usepackage{lipsum}
\usepackage{enumitem}
\setlist[enumerate]{itemsep=0mm}

\usepackage{JASA_manu}

\usepackage{threeparttable}
\usepackage{color, colortbl}

\usepackage[normalem]{ulem}
\usepackage{dutchcal}

\newcommand\redsout{\bgroup\markoverwith{\textcolor{red}{\rule[0.5ex]{2pt}{1pt}}}\ULon}

\definecolor{red}{rgb}{1,0,0}
\definecolor{blue}{rgb}{0,0,1}
\definecolor{green}{rgb}{0,0.6,0.4}

\definecolor{Gray}{gray}{0.9}

\newtheorem{theorem}{Theorem}

\makeatletter
\newcommand*{\centernot}{%
  \mathpalette\@centernot
}
\def\@centernot#1#2{%
  \mathrel{%
    \rlap{%
      \settowidth\dimen@{$\m@th#1{#2}$}%
      \kern.5\dimen@
      \settowidth\dimen@{$\m@th#1=$}%
      \kern-.5\dimen@
      $\m@th#1\not$%
    }%
    {#2}%
  }%
}
\makeatother

\newcommand{\blind}{1}

\begin{document}



\if1\blind
{
  \title{\bf Cost-Effectiveness Analysis for Disease Prevention -- A Case Study on Colorectal Cancer Screening}
  \author{Yi Xiong\\
    Department of Biostatistics, University at Buffalo, Buffalo, NY \\
        and \\
    Kwun C G Chan \\
    Department of Biostatistics, University of Washington, Seattle, WA \\
        and \\
    Malka Gorfine \\
    Department of Statistics and Operations Research, Tel Aviv \\University, Tel Aviv, Israel \\
    and \\
    Li Hsu \thanks{
    The authors gratefully acknowledge that this work is partially supported by the National Institutes of Health grants.}\hspace{.2cm}\\
    Biostatistics Program, Fred Hutchinson Cancer Center, Seattle, WA }
  \maketitle
} \fi

\if0\blind
{
  \bigskip
  \bigskip
  \bigskip
  \begin{center}
    {\LARGE\bf Cost-Effectiveness Analysis for Disease Prevention -- A Case Study on Colorectal Cancer Screening}
\end{center}
  \medskip
} \fi

\bigskip
\begin{abstract}

Cancer Screening has been widely recognized as an effective strategy for preventing the disease. Despite its effectiveness, determining when to start screening is complicated, because starting too early increases the number of screenings over lifetime and thus costs but starting too late may miss the cancer that could have been prevented. Therefore, to make an informed recommendation on the age to start screening, it is necessary to conduct cost-effectiveness analysis to assess the gain in life years relative to the cost of screenings. As more large-scale observational studies become accessible, there is growing interest in evaluating cost-effectiveness based on empirical evidence. In this paper, we propose a unified measure for evaluating cost-effectiveness and a causal analysis for the continuous intervention of screening initiation age, under the multi-state modeling with semi-competing risks. Extensive simulation results show that the proposed estimators perform well in realistic scenarios. We perform a cost-effectiveness analysis of the colorectal cancer screening, utilizing data from the large-scale Women's Health Initiative. Our analysis reveals that initiating screening at age 50 years yields the highest quality-adjusted life years with an acceptable incremental cost-effectiveness ratio compared to no screening, providing real-world evidence in support of screening recommendation for colorectal cancer.

\end{abstract}


\newpage


\section{Introduction}

Cancer preventive screening is recommended for various cancers in US and many countries \citep{ratushnyak2019cost}. 
As a preventive intervention, it targets people without cancer, unlike early-detection cancer screening, which aims to detect cancer at an early stage in individuals who already have cancer. Preventive screening focuses on identifying and removing precancerous lesions. Consequently, it is typical that a large number of people undergo screening, leading to potentially high costs per cancer prevented. Ideally, the benefits of screening should outweigh its costs \citep{khalili2020cost}. Our work is motivated by determining an optimal screening age for colorectal cancer (CRC). CRC is the third most common cancer and the second leading cause of cancer deaths worldwide \citep{sung2021global}. 
Yet it is also one of the most preventable cancers due to the effectiveness of endoscopy screening, which both detects and removes precancerous lesions such as polyps \citep{holme2017effectiveness, davidson2021screening}. Despite its effectiveness, determining when to start screening is complicated. Starting too early increases costs and risks associated with invasive procedures due to more screenings that an individual will have during their lifetime, and yet the cancer risk is generally low at young age. On the other hand, starting too late may miss the cancer that could have been prevented and as a result, have a devastating impact on an individual's quality of life and life expectancy. To provide informed recommendation on the optimal age to begin screening, it is necessary to conduct cost-effectiveness analysis to assess health gains  relative to the costs of starting screening at different ages. 

A common approach for the cost-effective analysis for evaluating cancer screening strategies is to use microsimulation models \citep{shen2006optimization, zauber2008evaluating, de2014benefits}, where a large number of individuals are simulated under some models that reflect the natural history of cancer development, and the benefits 
and costs of different screening strategies are evaluated. For example, microsimulation models for CRC use a structure that builds upon the adenoma-carcinoma sequence \citep{muto1975evolution}. Individuals start without polyps and, without screening intervention, may develop one or more adenomas, which can possibly develop into CRC, and then death from CRC or from competing risks that can occur at any time. Various screening strategies are then applied to these persons, and the benefit (measured by the number of quality-adjusted life-years gained) and costs (measured by the number of required colonoscopies) 
are quantified \citep{meester2018optimizing}. While microsimulation models are useful for comparing the effectiveness of different interventions, their underlying model structure and parameter values may not fully capture the complex disease process and the impact of screening on that process \citep{cronin1998assessing, thompson2004simulation}. 

There is substantial interest in examining cost-effectiveness based on empirical evidence. Our research aims to assess the cost-effectiveness of initiating CRC screening at different ages using data from the Women's Health Initiative (WHI) \citep{study1998design}. Launched in 1991, WHI is one of the largest U.S. disease prevention programs addressing various cancer types including CRC, heart diseases, and other  outcomes in postmenopausal women. The study enrolled about 100,000 women aged 50 to 79 years across US. It collected a wide range of sociodemographic and epidemiologic factors including well-curated CRC screening status over time. Given the relatively infrequent  occurrence of CRC, the extensive follow-up of WHI offers a unique opportunity to investigate the impact of the timing of CRC screening initiation on the risk of CRC and mortality over a lifetime. 

Methods for analyzing observational data need to address issues including censored outcomes, competing events, and confounding. There is a rich literature in cost-effectiveness analysis that accommodates censoring and adjusts for confounding factors. For example, \cite{chen2001causal} used a regression-based approach, assuming the Cox proportional hazards model \citep{cox1972regression}, to compare the restricted mean survival times between two groups. 
\cite{anstrom2001utilizing} employed the propensity score approach to estimate treatment effects on costs. \citet{li2018doubly} proposed a doubly-robust estimator to assess the effect of interventions based on both cost and effectiveness. However, most existing studies primarily focus on static intervention, whereas our research examines the cost-effectiveness of initiating screening at different ages. Although age is a continuous variable, our problem differs substantially from conventional continuous intervention settings, such as estimating dose-response functions \citep{imbens2000role}. First, the intervention is subject to censoring, as some individuals may die or develop the disease before starting screening, resulting in the intervention being censored. Second, since screening can prevent disease progression and prolong the disease-free period ultimately, reducing mortality, it is crucial to investigate its impact on disease development, in addition to mortality. 
Thus, it is essential to consider a multi-state framework with semi-competing risks, allowing us to examine disease and death jointly.  

Causal inference holds great significance and is embedded, both implicitly and  explicitly, in public health practice. Practitioners often base intervention decisions on presumed causal connections and the subsequent outcomes that arise from them \citep{glass2013causal}. In cost-effectiveness analysis, current methodologies for causal inference in healthcare decision-making are primarily derived from randomized trials \citep{willan2005regression,zhao2001estimating}. While randomized trials provide robust evidence for informing cost-effective interventions, observational data allows for analysis in more realistic settings. When dealing with a possibly censored continuous intervention such as the age at screening, larger sample sizes are necessary. For infrequent diseases like CRC, long follow-up periods are needed, which may be cost prohibitive. Observational studies like WHI offer readily available large-scale data for evaluating the cost-effectiveness of continuous interventions. The combination of observational data and causal inference can provide empirical evidence for informing  health policy or guidelines regarding the optimal timing of screening initiation. In the realm of causal inference, there have been a few studies examining the causal effects of the timing of treatment initiation or discontinuation \citep{cain2010start,hu2018modeling,yang2018modeling} and the related but different issue of truncation by death \citep{comment2019survivor,gao2020defining,nevo2022causal}. However, to our knowledge, no existing work has simultaneously addressed both the challenges of continuous intervention and multi-state modeling with semi-competing risks.


In this paper, we aim to conduct a causal cost-effectiveness analysis to determine the \textit{initial} screening age for CRC. The primary research question is: what would be the cost and benefit if screening were to begin at a specific age? To address this, we propose a unified measure to evaluate cost-effectiveness under a multi-state modeling framework. This comprehensive measure includes well-known quantities such as the restricted mean survival time and quality-adjusted life years \citep{royston2013restricted, zhao2016restricted, torrance1989utilities}, and it can also quantify the cost of intervention, such as the number of screenings during lifetime. We consider a multi-state process to account for multiple (dependent) event times that do not require the semi-Markov assumption often needed in other methods. We then develop a procedure to estimate the cost-effectiveness estimand using structural transition hazard models for the proposed multi-state process. 
The contributions of this paper are threefold. First, we introduce a comprehensive measure for both the effectiveness and cost of a possibly censored continuous intervention. Second, we extend the measure to multi-state modeling, providing novel insights into how cost-effectiveness is impacted by the initial screening age through disease development and various paths to death, with and without prior disease diagnosis. Third, we establish a clear definition of the causal estimand of interest and outline the necessary assumptions associated with the multi-state model for the possibly censored intervention and outcome. 

The rest of this paper is structured as follows. In Section 2, we describe the proposed cost-effectiveness measure and the causal inference framework, as well as the estimation procedure. Section 3 presents the cost-effectiveness analyses of WHI data regarding the initial screening age for CRC.  An extensive simulation study evaluating the performance of our proposed framework is shown in Section 4. Finally, Section 5 provides concluding remarks summarizing the key contributions of this research and potential future directions.

\section{Methodology}
\subsection{A Unified Measure for the Benefit and Cost} \label{sec:notation}
Consider an illness-death model (Figure 1), under which individuals can be in the healthy or diseased state during the follow-up. Let $K$ be the number of states, here, under the illness-death model, $K=2$. Further, let $T$ and $D$ be the age at disease onset and death, respectively. We define the benefit or cost at time $t$ by
\begin{eqnarray} \label{eqn.cost}
M(t) & = & \sum_{k=1}^K \int_0^t \widetilde{Y}_k(u) dW_k(u),\vspace{-7mm}
\end{eqnarray}
where $\widetilde{Y}_k(\cdot)$ is an indicator function, which is 1 if an individual in the $k$th state and alive (i.e., $D \geq u$) and 0 otherwise, and $W_k(\cdot)$ is a cumulative weight function associated with the $k$th state, $k=1, \ldots, K$. $M(t)$ encompasses many measures for quantifying benefit or cost. Below we show several examples for $M(t)$ and provide more details in Section \ref{sec:example}. 
\begin{figure}[h!]
    \centering  \includegraphics[scale=0.38]{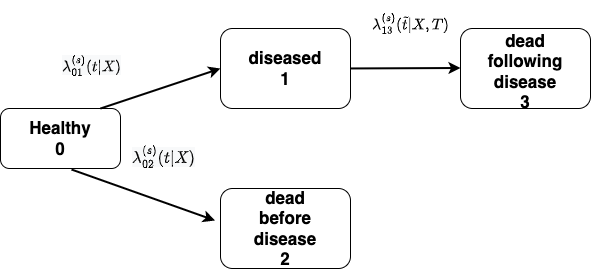}
    \caption{An illness-death model} \label{fig:multistate}
\end{figure}
\vspace{-5pt}
\begin{enumerate}
    \item[] \textit{Example 1: Restricted Mean Survival Time (RMST).} The RMST is a popular measure for quantifying the average survival experience up to a certain time point, $t$ \citep{zhao2016restricted}. Mathematically, it can be defined as the area under the survival curve up to time $t$. In this case, there is only one event of interest, death. 
    Setting $W_k(u) = u, k = 1, \ldots, K$, we have $E\{M(t)\}  = E\{\int_0^t \sum_{k=1}^K \widetilde{Y}_k(u) du\} = 
    \int_0^t \Pr(D \geq u) du$, which represents the RMST. We can also set $K=1$ to obtain RMST, but we opt for $K=2$ to more effectively capture the impact of screening on the disease process rather than solely modeling overall mortality. 
    \item[] \textit{Example 2: Cause-Specific Mean Life Years Lost.} Life years lost is complementary to the RMST, and it offers an appealing attribute to decompose the years lost due to specific cause pathways leading to mortality before time $t$ \citep{andersen2013decomposition}. Define the expected life years lost $E\{L(t)\}  =  t - E\{M(t)\}$. 
    Under the illness-death model, an individual may die before the disease occurs, as indicated by the path $0 \rightarrow 2$ or first develop the disease and then die, as indicated by the path $0 \rightarrow 1 \rightarrow 3$. Let $W_k(t) = t$, for $k=1$ and $2$, and $f_D(\cdot|T,X)$ be the conditional probability density function of $D$ given $T$ and $X$,
    the mean life years lost can then be decomposed into 
    \begin{eqnarray*}
 E\{L(t)\} 
   & = &\int_{0}^t \int_0^ u f_D(v| T\geq v, X) \Pr(T\geq v|X) dv du  \\
   & & \mbox{} + \int_{0}^t \int_0^u f_D(v|T<v , X) \Pr(T<v|X) dv du. \vspace{-7mm}
\end{eqnarray*} 
The first term is the mean life years lost through the path $0 \rightarrow 2$ and second term is the mean life years lost through the path $0 \rightarrow 1 \rightarrow 3$. 
This is particularly informative when assessing the performance of an intervention. In the CRC example, colonoscopy has been shown to be effective in reducing CRC incidence rates substantially. However, CRC is still relatively infrequent in a population, despite it is a common cancer. As a result, the gain in life years in the population may not be apparent to reflect the effectiveness of colonoscopy because the cancer only affects a small fraction of the population. By examining cause-specific mean life years lost, it can shed a clearer light on the cost-effectiveness of the intervention in relation to the disease that it targets.  
\item[] \textit{Example 3: Quality-Adjusted Life Years (QALY). } In Example 1, $E\{M(t)\}$ represents the expected life years up to time $t$ when the time origin is set at birth. However, the weight function $W_k(u)$ can be set to reflect changing quality score over time in the $k$th state. In this situation, $E\{M(t)\}$ becomes the expected QALY up to time $t$. The cause-specific quality-adjusted mean life years lost can be defined accordingly. 
\item[] \textit{Example 4: Number of Screenings.} When $W_1(u)$ is the cumulative number of screenings up to time $u$ and $W_2(u) = 0$, then $E\{M(t)\}$ measures the expected number of screenings prior to disease or death, whichever comes first.
\end{enumerate}
Having a unified measure for the benefit or cost forms a fundamental basis for evaluating the cost-effectiveness of an intervention. Decision-makers are often faced with the challenge of determining the value of a particular intervention relative to its associated costs. This becomes especially pertinent when resources are limited, and choices must be made about how to optimize these resources to achieve the best outcomes. By establishing a unified measure, it allows us to assess both the benefit and cost or the combination of benefit and cost such as incremental cost effectiveness ratio (incremental cost/incremental benefit) of different interventions systematically.

\subsection{Causal Inference of a Continuous Intervention under the Multi-State Model}
Our interest lies in assessing the effects of different initiation timings for screening on both the benefit (e.g., increased survival years) and cost (e.g., number of screenings) from a cost-effectiveness standpoint.  
Because the screening age for an individual in an observational study cohort is continuous, 
we cannot directly adopt the conventional counterfactual framework for a binary intervention (e.g., treatment vs. control). Instead, we consider this to be a counterfactual problem with a continuous intervention. However, as mentioned above, age at which intervention is initiated is subject to censoring, creating additional challenges.  The question that we ask is: what would the benefit and cost be if a  screening policy recommended screening starting at age $s$? 

Let $T^*$ and $D^*$ denote the potential age at disease onset and potential age at death had the individual never undergone the screening. For $s<\min(T^*,D^*)$, we define $T^{(s)}$ and $D^{(s)}$ as the potential age at disease onset and age at death, respectively, if screening had occurred at $s$. If an individual had not undergone screening prior to the disease onset or death, i.e., $s\geq\min (T^*,D^*)$, then $T^{(s)}=T^*$ and $D^{(s)}=D^*$. Define $X$ to be a vector of covariates measured at baseline. 
As shown in Figure \ref{fig:multistate} for the illness-death model, individuals start in state 0 (e.g., healthy) and then either move to state 2 (e.g., death without disease) directly, or transit first to state 1 (e.g., disease) and then to state 3 (e.g., death following disease incidence). Given the continuous screening time $s$, the illness-death model is defined by three hazard functions in terms of potential outcomes $T^{(s)}$ and $D^{(s)}$: the hazard functions from healthy to disease and death, respectively, as well as the hazard function from disease to death. Specifically, the hazards functions from healthy to disease and death are defined under the competing risks framework as following: 
\begin{align}
      \lambda^{(s)}_{01}(t|X) & =\lim_{\Delta t\rightarrow 0} \Delta t^{-1}\Pr(t \leq T^{(s)} < t+\Delta t | T^{(s)} \geq t, D^{(s)} \geq t, X),  \label{hzd01.pot} \\
     \lambda^{(s)}_{02}(t|X) & =\lim_{\Delta t\rightarrow 0} \Delta t^{-1}\Pr(t \leq D^{(s)} < t+\Delta t | T^{(s)} \geq t, D^{(s)} \geq t, X), \label{hzd02.pot} 
\end{align}
for $t>0$. 
For the hazard function from disease to death, we define it based on the sojourn time, $\widetilde D^{(s)} = D^{(s)} - T^{(s)}$. The commonly used semi-Markov assumption implies that $\widetilde D^{(s)}$ is independent of $T^{(s)}$ given risk factors $X$. However, this assumption may not be realistic, as for many diseases the survival is associated with age at disease, see e.g., \cite{van2012association}. 
To account for such possible dependence, we define the hazard function including age at disease as a covariate,
\begin{align}
      \lambda^{(s)}_{13}(\tilde{t}|X,T^{(s)}) & = \lim_{\Delta \tilde{t}\rightarrow 0} (\Delta \tilde{t})^{-1}\Pr(\tilde{t} \leq \widetilde D^{(s)} < \tilde{t}+\Delta \tilde{t} | \widetilde D^{(s)} \geq \tilde{t}, T^{(s)}, X), \label{hzd13.pot} 
\end{align}
for $\tilde{t} > 0$. Here we use a different notation  $\tilde{t}$ to indicate that it represents a sojourn time starting at the age of disease onset
\citep{andersen2000multi}.

Both potential time to disease onset and time to death are subject to right censoring time $C$. We assume that $C$ is independent of both times given the risk factor $X$, as stated below.  
\begin{itemize}
      \item[] {\sc Assumption A1} (Conditional Independence of Censoring). {\it The censoring time $C$ is independent of $\{T^*, D^*, T^{(s)},D^{(s)}, 0<s \leq \min(T^*, D^*) \}$ given the risk factor $X$.}
\end{itemize}
The conditional independence assumption (A1) also implies that given $T^{(s)}$ and $X$, the sojourn time $\tilde{D}^{(s)}=D^{(s)}-T^{(s)}$ is independent of $C-T^{(s)}$. This allows us to estimate $\lambda_{13}^{(s)}(\tilde{t}|X, T^{(s)})$ in \eqref{hzd13.pot} by e.g., including $T^{(s)}$ as a covariate, without requiring additional assumptions about the censoring mechanism. 
Now, define the potential observational time as $U^{(s)}=\min(T^{(s)},D^{(s)},C)$. This represents the potential age at disease onset if the disease indicator $\delta_1^{(s)} \equiv I(T^{(s)} \leq \min(D^{(s)},C) )$ equals 1, the potential age at death if the death indicator $\delta_2^{(s)} \equiv (1-\delta_1^{(s)})I(D^{(s)} \leq C )$ equals 1, and the time at last observation, i.e., the censoring time $C$, if both $\delta_1^{(s)} $ and $\delta_2^{(s)}$ are 0. 
For the rare occasions when the disease is diagnosed at the time of death, we assume $T^{(s)}$ occurs before $D^{(s)}$. 
Further, define $V^{(s)}=\delta_{1}^{(s)} \{\min(D^{(s)},C) - T^{(s)}\}$, which is the potential time to death since the potential disease occurrence if $\delta_{3}^{(s)}\equiv\delta_{1}^{(s)} I(D^{(s)}\leq C)$ equals 1, and time to last observation since the potential disease occurrence, otherwise. In addition, we define the following potential outcome counting processes, 
\begin{align*}
    N^{(s)}_{01}(t) & = I(U^{(s)}\leq t, \delta_1^{(s)}=1),
    N^{(s)}_{02}(t) = I(U^{(s)}\leq t, \delta_2^{(s)}=1), 
    N^{(s)}_{13}(\tilde{t})  = I(V^{(s)}\leq \tilde{t}, \delta_3^{(s)}=1),\nonumber \\
     Y^{(s)}_{0}(t) & = I(U^{(s)}\geq t),
     Y^{(s)}_{1}(\tilde{t}) = I(V^{(s)}\geq \tilde{t}).\nonumber
\end{align*} 

\noindent Note that the time-shifted counting process $N^{(s)}_{13}(\tilde{t})$ denotes the potential outcome death up to the sojourn time $\tilde{t}$ from the disease onset when screening occurs at time $s$. Similarly, $Y^{(s)}_{1}(\tilde{t})$ indicates whether an individual is potentially at risk in the disease state at the sojourn time $\tilde{t}$.
Under Assumption A1, the potential outcome hazard function can be represented in terms of event counting and at-risk processes. Let $dN_{0k}^{(s)}(t) = N_{0k}^{(s)}(t^- + \Delta t) - N_{0k}^{(s)}(t^-)$, $k = 1, 2$, and $dN_{13}^{(s)}(\tilde t) = N_{13}^{(s)}(\tilde t^- + \Delta \tilde t) - N_{0k}^{(s)}(\tilde t^-)$, Eq. (\ref{hzd01.pot}) -- (\ref{hzd13.pot}) can be written as 
 \begin{align*}
         \lambda^{(s)}_{0k}(t|X) & = \lim_{\Delta t \rightarrow 0} \frac{1}{\Delta t} \Pr\{ dN_{0k}^{(s)}(t) = 1| Y_0^{(s)}(t), X \}  \\
         \lambda^{(s)}_{13}(\tilde{t}|T^{(s)},X) & = \lim_{\Delta \tilde{t} \rightarrow 0} \frac{1}{\Delta \tilde{t}} \Pr\{ dN_{13}^{(s)}(\tilde{t}) = 1 | Y_1^{(s)}(\tilde{t}),T^{(s)}, X \}  \, .
          \end{align*}

Now, we define the observed data. Let $U=\min(T,D,C)$ and $V=\delta_1\{\min(D,C)-T\}$ denote the observed time to the first event of disease or death and the observed time to death since disease, respectively, where $\delta_1=I(T\leq \min(D,C))$ indicates disease status. Let $\delta_2=(1-\delta_1)I(D\leq C)$, and $\delta_3=\delta_1I(D\leq C)$ be the indicators of death without disease and death after disease, respectively. The observed counting processes are given by
\begin{align*}
    & N_{01}(t)  = I(U \leq t, \delta_{1} = 1), 
    N_{02}(t)  = I(U \leq t, \delta_{2} = 1), 
    N_{13}(\tilde{t})  = I(V\leq \tilde{t}, \delta_{3} = 1), 
\nonumber\\
    & Y_{0}(t)  = I(U \geq t), 
     Y_{1}(\tilde{t})  = I(V\geq \tilde{t}),\nonumber
\end{align*}
where time $\tau$ is the end of a study or a pre-specified time point and $\Pr(Y_0(\tau) Y_1(\tau)>0) > 0$. Further let $S$ denote the time at which the individual initiates screening. The observational screening history up to age $U$ is denoted by $\bar{Z}(U;S) = \{Z(u;S)=I(S < u), 0<u\leq U \}$.

The following consistency assumption links the potential outcomes to the observed processes.
\begin{itemize}
    \item[] {\sc Assumption A2} (Consistency). {\it For individuals who have initiated screening prior to $U$, i.e. $S<U$, $N_{01}(t) = N^{(S)}_{01}(t), N_{02}(t) = N^{(S)}_{02}(t)$, and $N_{13}(\tilde{t}) = N^{(S)}_{13}(\tilde{t})$ as well as the at-risk processes $Y_{0}(t) = Y^{(S)}_{0}(t)$ and $Y_{1}(\tilde{t}) = Y^{(S)}_{1}(\tilde{t})$; for individuals who have not undergone screening prior to $U$,  $N_{01}(t) = N^{(s)}_{01}(t), N_{02}(t) = N^{(s)}_{02}(t)$, $N_{13}(\tilde{t}) = N^{(s)}_{13}(\tilde{t})$, $Y_{0}(t) = Y^{(s)}_{0}(t)$, and $Y_{1}(\tilde{t}) = Y^{(s)}_{1}(\tilde{t})$
    where $s\geq U$, for all $t>0$ and $\tilde{t} > 0$. }
\end{itemize}
This consistency assumption suggests that since an individual who has not undergone screening prior to $U$ is consistent with any regime $s$, $s \geq U$, his/her observed processes equal the corresponding potential processes if this individual followed regime $s$. In the absence of censoring, the potential processes would be for $T^*$ and $D^*$ had the individual never undergone the screening. 
In addition, the following two identifiability assumptions are required.
\begin{itemize}
     \item[] {\sc Assumption A3} (No Unmeasured Confounding). {\it Conditional on baseline risk factors $X$, the potential outcome hazard functions of disease and mortality at time $t$ with screening time $s$ are independent of the observed screening history up to and including time $t$. In other words given Assumptions A1 and A3, }
     \begin{align*}
         \lambda^{(s)}_{0k}(t|X) 
         & = \lim_{\Delta t \rightarrow 0} \frac{1}{\Delta t} \Pr \{ dN_{0k}^{(s)}(t) = 1| Y_0^{(s)}(t), X, \bar{Z}(t;S) \}, 
         \quad \mbox{for } k = 1, 2,       \\
         \lambda^{(s)}_{13}(\tilde{t}|T^{(s)},X) 
         & = \lim_{\Delta \tilde{t} \rightarrow 0} \frac{1}{\Delta \tilde{t}} \Pr(dN_{13}^{(s)}(\tilde{t}) = 1 | Y_1^{(s)}(\tilde{t}),T^{(s)}, X, \bar{Z}(T^{(s)};S)).
          \end{align*}
     \item[] {\sc Assumption A4} (Positivity).
     {\it There exists a positive constant $\epsilon > 0$ such that the hazard rate function of observed screening at time $t$ given $X$ is bounded below by $\epsilon$ and finite, i.e.,
     $\epsilon < \lim_{\Delta t\to 0}P(t\leq S < t+\Delta t|U\geq t, S \geq t, X)\big/\Delta t < \infty$. }
\end{itemize}

Assumption A3 implies that the baseline risk factors are sufficient to account for the self selected screening time in the observational data. Along with Assumption A2, it establishes a connection between the causal estimands to the observed data. This connection facilitates the inference of the causal effect of the timing of screening initiation. While $X$ in this work consists of only baseline risk factors, there could be some relevant time-dependent risk factors (e.g., body mass index). The assumption can be further relaxed by allowing for time-dependent risk factors up to time $t$. However, when predicting the risk and determining the appropriate timing to initiate screening, for practical reasons, we opt to assume that the decision is made based on the risk factors observed at baseline, rather than considering potential risk factor values that may develop in the future.  
Assumption A4 implies that every individual has a non-zero probability of undergoing screening at any age prior to the occurrence of disease or death. Given that our primary focus is on the preventive effect of screening, we consider the time to screening for individuals who have not undergone screening before the onset of the disease to be censored by the time of disease onset.  

To evaluate the cost-effectiveness concerning the recommended screening age $s$, our interest lies in the following function,
\begin{eqnarray}
 E\{M^{(s)}(t)\} & = &  E_X \left [ \int_{0}^t E\{ I(T^{(s)}\geq u,D^{(s)}\geq u)|X\}dW_1(u) \right ] \nonumber \\
 \quad & & \mbox{} +E_X \left [ \int_{0}^t E\{ I(T^{(s)}\leq u,D^{(s)}\geq u)|X\}dW_2(u) \right ]. \label{eq:costeffectExpect3}
\end{eqnarray} 
Let $\Lambda_{0k}^{(s)}(t) = \int_0^t \lambda_{0k}^{(s)}(u) du$, $\Lambda_{0k}(t) = \int_0^t \lambda_{0k}(u) du$, $k=1, 2$, $\Lambda_{13}^{(s)}(t) = \int_0^t \lambda_{13}^{(s)}(u) du$, and $\Lambda_{13}(t) = \int_0^t \lambda_{13}(u) du$, the first term on the right hand side of \eqref{eq:costeffectExpect3} equals
\begin{eqnarray}
\lefteqn{E_X \left [ \int_0^t \exp\{-\Lambda_{01}^{(s)}(u|X) - \Lambda_{02}^{(s)}(u|X)\} dW_1(u) \right ]} \nonumber\\ 
& =  & E_X \left [ \int_0^t \exp\{-\Lambda_{01}^{(s)}(u|X, \bar{Z}(u;S)) - \Lambda_{02}^{(s)}(u|X, \bar{Z}(u;S))\} dW_1(u) \right ], \nonumber
\\ 
& =  & E_X \left [ \int_0^t \exp\{-\Lambda_{01} (u|X, \bar{Z}(u;S=s)) - \Lambda_{02}(u|X, \bar{Z}(u;S=s))\} dW_1(u) \right ], \nonumber\\ 
& \equiv &  E_X \left \{ \int_0^t \Pr(\min(T, D) \geq u | X, \bar{Z}(u;S=s)) dW_1(u) \right \}, \label{surv1}
\end{eqnarray}
where the first equality is due to Assumption A3 and the second equality is due to Assumption A2. 
Following similar arguments, the second term on the right hand side of \eqref{eq:costeffectExpect3} equals
\begin{eqnarray}
\lefteqn{E_X \left [ \int_0^t\int_0^u \Pr(D^{(s)}\geq u|T^{(s)}=v,X)dF_{T^{(s)}}(v|X)dW_2(u) \right ] } \nonumber \\
& =  & E_X \left [ \int_0^t\int_0^u \exp\{-\Lambda_{13}^{(s)}(u-v|T^{(s)}=v,X)\}dF_{T^{(s)}}(v|X)dW_2(u)\right ] \nonumber \\
& =  & E_X\left [  \int_0^t\int_0^u \exp\{-\Lambda_{13}^{(s)}(u-v|T^{(s)}=v,X,\bar{Z}(v;S))\}dF_{T^{(s)}}(v|X,\bar{Z}(v;S))dW_2(u)\right ] \nonumber
\\
& =  & E_X \left [ \int_0^t\int_0^u \exp\{-\Lambda_{13}(u-v|T=v,X,\bar{Z}(v;S=s))\}dF_{T}(v|X,\bar{Z}(v;S=s))dW_2(u) \right ],\nonumber \\
& \equiv  & E_X \left \{ \int_0^t  \Pr(T< u, D \geq u | X, \bar{Z}(u;S=s))  dW_2(u) \right \}, \label{surv2}
\end{eqnarray}
where $F_{T^{(s)}}(\cdot)$ is the cumulative incidence function (CIF) for the potential time at disease onset, defined as
$F_{T^{(s)}}(u|X) = \int_0^u\exp\{-\Lambda_{01}^{(s)}(v|X) - \Lambda_{02}^{(s)}(v|X)\}d\Lambda_{01}^{(s)}(v|X)$ = $ \int_0^u\exp\{-\Lambda_{01}(v|X,\bar{Z}(v;S=s))-\Lambda_{02}(v|X,\bar{Z}(v;S=s))\}d\Lambda_{01}(v|X,\bar{Z}(v;S=s))$, and the second and third equalites are due to Assumptions A3 and A2, respectively.

Therefore, to estimate the causal estimand $E\{M^{(s)}(t)\}$ for a given screening age $s$, it has come down to estimate the hazard functions $\lambda_{0k}(t|X,\bar{Z}(v;S=s))$, $k = 1, 2$, and $\lambda_{13}(t|X,\bar{Z}(v;S=s))$ with observed screening history $\bar{Z}(v;S=s)$ and baseline risk factor $X$. 
We employ the Cox proportional hazards models 
\begin{eqnarray}
\lambda_{0k}(t|\bar{Z}(t;S),X) & = & \lambda_{0k,0}(t)\exp(\beta_{0k}Z(t;S)+\gamma'_{0k}X), \qquad k = 1, 2, \label{eq:multimodels1}\\
\lambda_{13}(\tilde{t}|T,\bar{Z}(T;S),X) & = & \lambda_{13,0}(\tilde{t})\exp(\alpha T+\beta_{13}Z(T;S)+\gamma'_{13}X),\label{eq:multimodels3}
\end{eqnarray}
where $\lambda_{01,0}(t)$, $\lambda_{02,0}(t)$ and $\lambda_{13,0}(\tilde t)$ are unspecified baseline hazard functions, and $\theta_{01} = (\beta_{01},\gamma_{01}')'$, $\theta_{02} = (\beta_{02},\gamma_{02}')'$ and $\theta_{13} = (\alpha, \beta_{13}, \gamma_{13}')'$ are the regression coefficients.


\subsection{Estimation} \label{sec:est}
The observed data consist of $n$ independently and identically distributed random variables, $\mathcal{O}_i, i = 1, \ldots, n$, where $\mathcal{O}_i=\{U_i,V_i, \delta_{1i},\delta_{2i}, \delta_{3i},  \bar{Z}(U_i; S_i) \}$. The estimation procedure  under the Cox proportional hazards model with competing risks and time-dependent covariates has been well established; see e.g., \cite{kalbfleisch2011statistical}. Hence, the maximum partial likelihood estimators $\widehat \theta_{01}$ and $\widehat \theta_{02}$ can be obtained by solving the following respective partial likelihood score equations
\begin{eqnarray*}
\sum_{i=1}^n \int_0^\tau \left \{ \widetilde{Z}_i(t)  - \frac{s_{01}^{(1)}(t; \theta_{01})}{s_{01}^{(0)}(t; \theta_{01})} \right \} N_{i01}(dt) & = & 0,\vspace{-5mm}
\end{eqnarray*}
and
\vspace{-3mm}
\begin{eqnarray*}
\sum_{i=1}^n \int_0^\tau \left \{ \widetilde{Z}_i(t)  - \frac{s_{02}^{(1)}(t; \theta_{02})}{s_{02}^{(0)}(t; \theta_{02})} \right \} N_{i02}(dt) & = & 0,
\end{eqnarray*}
where $N_{i01}(t), N_{i02}(t), Y_{i0}(t)$ are the realizations of $N_{01}(t), N_{02}(t)$ and $Y_{0}(t)$ for $i$th individual, and $\widetilde{Z}_i(t)' = (Z(t;S_i), X_i')$,  $s_{01}^{(j)}(t; \theta_{01}) = \sum_{i=1}^n Y_{i0}(t) \widetilde{Z}_i^{\otimes j}(t;S_i) \exp\{\beta_{01}Z(t;S_i)+\gamma'_{01}X_i\}$, and \\
$s_{02}^{(j)}(t; \theta_{02}) = \sum_{i=1}^n Y_{i0}(t) \widetilde{Z}_i^{\otimes j}(t) \exp\{\beta_{02}Z(t;S_i)+\gamma'_{02}X_i\}$ with $Z^{\otimes j} = 1$ and $Z$ for $j = 0$ and $1$, respectively. 
Under the competing risks framework, the dependence between time to disease and time to death is left unspecified and they are not assumed to be independent \citep{kalbfleisch2011statistical}. While the data are presented as right censoring, the estimation procedure can straightforwardly accommodate left truncation data by appropriately adjusting the risk set $Y_{i0}(t) = I(U_i \geq t > L_i)$ where $L_i$ is the left truncation time for the $i$th individual.

Let $N_{i13}(\tilde{t})$ and $Y_{1i}(\tilde t)$ represent realizations of the time-shifted counting processes $N_{13}(\tilde{t})$ and $Y_{1}(\tilde t)$ for the $i$th individual since disease onset. The time origin has shifted from study entry to the onset of the disease, and the partial likelihood now includes only individuals who have developed the disease. We can then construct a partial likelihood function for the parameters $\theta_{13}$ in the model \eqref{eq:multimodels3} given the data $N_{i13}(\tilde{t})$ and $Y_{1i}(\tilde t)$ for all individuals who develop the disease. Notably, this likelihood yields a consistent estimator $\widehat\theta_{13}$, because the model \eqref{eq:multimodels3} includes $T$ and $X$ as covariates. Under Assumption A1, it is implied that the sojourn time $(D-T)$ and the (sojourn) censoring time $(C-T)$ are independent, given $T$ and $X$. Consequently, we can obtain the maximum partial likelihood estimator $\widehat\theta_{13}$  by solving the score equation 
\begin{eqnarray*}
\sum_{i=1}^n \int_0^{\tau} \left \{ \widetilde{Z}_{i1}  - \frac{s_{13}^{(1)}(\tilde{t}; \theta_{13})}{s_{13}^{(0)}(\tilde{t}; \theta_{13})} \right \} N_{i13}(d\tilde{t}) & = & 0,
\end{eqnarray*}
where $\widetilde{Z}'_{i1}= (T_i, Z(T_i;S_i), X_i')$,  $s_{13}^{(j)}(\tilde{t}; \theta_{13}) = \sum_{i=1}^n Y_{i1}(\tilde{t}) \widetilde{Z}_{i1}^{\otimes j} \exp\{\alpha T_i + \beta_{13}Z(T_i;S_i)+\gamma'_{13}X_i\}$. 
The cumulative baseline hazard functions $\{\Lambda_{01,0}(t)$, $\Lambda_{02,0}(t)$, and $\Lambda_{13,0}(t)\}$ can be estimated with the Breslow estimators provided in Supplemental Materials (SM) Section 1. 

An estimator of $E\{M^{(s)}(t)\}$ can then be obtained by plugging $\{\widehat \theta_{01}, \widehat \theta_{02}, \widehat\theta_{13} \}$ and the Breslow estimators  $\{\widehat \Lambda_{01,0}(t), \widehat \Lambda_{02,0}(t), \widehat \Lambda_{13,0}(t)\}$ into Eqs. (\ref{surv1}) and (\ref{surv2}). Specifically, denote $P_1(t|X,s)  = \Pr(\min(T,D)\geq t|\bar{Z}(t;S=s),X)$, $P_{13}(t|r, X, s) = {P}(D \geq  t|T=r,\bar{Z}(T;S=s),X)$ and $P_2(t|X,s)  = \Pr(D\geq t,T\leq t|\bar{Z}(t;S=s),X)$, and 
the estimators are 
\begin{eqnarray*}
    \widehat{P}_1(t|X, s) & = & \exp \left \{ - \widehat{\Lambda}_{01}(t|\bar{Z}(u;S=s),X) - \widehat{\Lambda}_{02}(t|\bar{Z}(u;S=s),X) \right \}, \\
    \widehat{P}_{13}(t|r, X, s) &=  & \exp\left\{-\widehat{\Lambda}_{13}(t-r|T=r,\bar{Z}(r;S=s),X) \right\}, \\
  \widehat{P}_2(t|X, s )  & = & \int_{0}^t\widehat{P}_{13}(t|r, X, s) d\widehat{F}_{T}(r|\bar{Z}(r;S=s),X), 
\end{eqnarray*}
where cumulative hazard function estimators $\widehat\Lambda_{0k}(t|\bar{Z}(t; S=s, X) = \int_0^t \exp\{\widehat\beta_{0k}Z(u; S=s) + \widehat \gamma_{0k}' X\} \widehat \Lambda_{0k,0} (du)$ for $k =1, 2$ and  $ \widehat{\Lambda}_{13}(t-r|T=r,\bar{Z}(r;S=s),X) = \int_0^{t-r} \exp\{\widehat\alpha r +\widehat \beta_{13}Z(r;S=s)+\widehat \gamma'_{13}X\} \widehat\Lambda_{13,0}(du) $, as well as cumulative disease incidence estimator $\widehat F_T(t|\bar{Z}(t;S=s),X) =\int_0^t  \widehat{P}_1(u|X, s) \exp\{\widehat{\beta}_{01}Z(u;S=s)+\widehat{\gamma}'_{01}X\}\widehat{\Lambda}_{01,0}(du) $. The estimator of $E\{M^{(s)}(t)\}$ is thus given by
\begin{align}
\label{eq:estM1}
     \widehat{E}\{M^{(s)}(t)\}=\frac{1}{n}\sum_{i=1}^n\left(\int_{0}^t  \widehat{P}_1(u|X_i, s) dW_1(u)+\int_{0}^t  \widehat{P}_2(u|X_i, s ) dW_2(u) \right).
\end{align}
To obtain the variance estimator, we draw $B$ bootstrap samples from the observed data $\{\mathcal{O}_i, i=1,\ldots,n\}$ and obtain $\widehat{E}\{M^{(s)}(t)\}$ with each bootstrap sample. The variance of $\widehat{E}\{M^{(s)}(t)\}$ is estimated by the empirical variance over $B$ bootstrap samples. We have the following theorem:
\begin{theorem}
\textit{Under Assumptions (A1)--(A4) and regularity conditions (B1)--(B5), for any given $s\in [0,t]$, as $n$ goes to infinity, }
\begin{enumerate}
    \item $ \widehat{E}\{M^{(s)}(t)\}$ converges to $E\{M^{(s)}(t)\}$ uniformly over $t\in[0,\tau]$.
    \item $\sqrt{n}[ \widehat{E}\{M^{(s)}(t)\} - E\{M^{(s)}(t)\}]$ converges weakly to a mean-zero Gaussian process over $t\in[0,\tau]$.
    \item The bootstrap-based variance estimator for $ \widehat{E}\{M^{(s)}(t)\}$ is a consistent estimator of the asymptotic variance of $ \widehat{E}\{M^{(s)}(t)\}$.
\end{enumerate}
\end{theorem}
The main step of the proof involves demonstrating that Eq. (\ref{eq:estM1}) asymptotically equals a sum of independent and identically distributed martingales. The subsequent part of the proof utilizes martingale theory. The details of the proof are provided in SM Section 2. 
\subsection{Some Common and Useful Estimands} \label{sec:example}
In this section, we provide a detailed derivation of examples of $E\{ M^{(s)}(t)\}$ as outlined in Section~\ref{sec:notation} under the illness-death model (Figure~\ref{fig:multistate}). 
\subsubsection{Example 1: RMST}
The causal estimand for the RMST can be obtained by assuming  $W_1(u) = W_2(u) = u$ in Eq. (\ref{eq:costeffectExpect3}) for $E\{M^{(s)}(t)\}$. Combining Eqs. (\ref{surv1}) and (\ref{surv2}) yields the RMST estimand $E\{M^{(s)}(t)\} = 
E_X \{ \int_{0}^t E\{ I(D^{(s)}\geq u)|X\}du\}$, which equals $
\int_0^t \Pr(D \geq u |\bar{Z}(u;S=s),X)du$. Since it solely concerns mortality, it is sufficient to model only time to death, $D$, which has been the conventional approach \citep{zhao2016restricted}. Here, we decompose the overall survival probability into two components to account for the disease an individual could experience. This allows for a better modeling of the screening effect on the disease process, and subsequently, death. Therefore, the RMST can be written as
$E\{M^{(s)}(t)\} =  E_{X} \left \{ \int_0^t P_1(u|X,s)  du \right \} + E_{X} \left \{ \int_0^t P_2(u|X,s) du \right \}$, which can be estimated by \vspace{-5mm}
\begin{eqnarray*} \label{eq:rmstest}
    \widehat{E}\{M^{(s)}(t)\} & = & \frac{1}{n}\sum_{i=1}^n\left(\int_{0}^t \widehat{P}_1(u|X_i,s)du+\int_{0}^t \widehat{P}_2(u|X_i,s)du\right).
\end{eqnarray*}

\subsubsection{Example 2: Cause-Specific Mean Life Years Lost}

We define the estimand for the mean life years lost given recommended screening at age $s$ to be  $ E\{L^{(s)}(t)\}  =  t - E\{M^{(s)}(t)\} $. It can be further decomposed into sum of two cause-specific estimands: (1) mean life years lost through the disease pathway $E\{L^{(s)}_{0 \rightarrow 1 \rightarrow 3} (t)\} = E_X \{\int_{0}^t \int_0^ u f_{ D^{(s)}}(v|T^{(s)}<v, X)  \Pr(T^{(s)}<v|X) dv du\} $; and (2) mean life years lost due to other causes $ E\{L^{(s)}_{0 \rightarrow 2} (t)\} =  
E_X \{ \int_{0}^t \int_0^ u f_{ D^{(s)}}(v| T^{(s)}\geq v, X) \Pr( T^{(s)}\geq v|X )dv du \}$, where $f_{D^{(s)}}(v|T^{(s)}, X)$ is the conditional probability density function of $D^{(s)}$ given $T^{(s)}$ and $X$.
After some algebra and under Assumptions A2 and A3 (details are provided in SM Section 3), these two terms can be written as following 
\begin{eqnarray*}
E\{L^{(s)}_{0 \rightarrow 1 \rightarrow 3}(t)\}  & = &  E_X \left \{\int_{0}^t \int_0^ u \{1-P_{13}(u|v,X,s)\}   P_1 (v|X,s) \Lambda_{01}(dv|X, {Z}(v;S=s)) du \right \},
\end{eqnarray*} 
\vspace{-5mm}
\noindent and
\vspace{-5mm}
\begin{eqnarray*}
E\{L^{(s)}_{0 \rightarrow 2} (t)\} & = & E_X \left \{\int_{0}^t \int_0^ u P_1 (v|X,s) \Lambda_{02}(dv|X, {Z}(v;S=s))du \right \}.
\end{eqnarray*} 
Therefore, the expected loss years due to disease ($0 \rightarrow 1 \rightarrow 3$) and other causes ($0 \rightarrow 2$) before time $t$ for individuals with  screening at age $s$ can be respectively estimated by
 $\widehat{E}\{L^{(s)}_{0 \rightarrow 1 \rightarrow 3} (t)\}   =   \frac{1}{n}\sum_{i=1}^n \int_0^t\int_0^u \{ 1-\widehat{P}_{13}(u|v,X_i,s) \} \widehat{P}_1(v|X_i,s) d\widehat{\Lambda}_{01}(v|Z(v;S_i=s),X_i)du$, 
and 
 $\widehat{E}\{L^{(s)}_{0 \rightarrow 2} (t)\} = $ \\ $  \frac{1}{n}\sum_{i=1}^n \int_0^t\int_0^u \widehat{P}_1(v|X_i,s)d\widehat{\Lambda}_{02}(v|Z(v;S_i=s),X_i)du$. 

\subsubsection{Example 3: QALY and Quality-Adjusted Years Lost}
Similar to the estimand for the RMST, we can define the estimand for the QALY. Let $Q(u)$ denote a quality score function ranging from 0 to 1 with 1 being healthy and 0 being death, and set $W(u) = \int_0^{u} Q(v)dv=u$ to indicate the cumulative quality score up to time $u$. Under the illness-death model where an individual is healthy, $Q(u) = 1$. If an individual is in the diseased state, the quality score
will be below 1. Following the unified cost-effectiveness measure $E\{M^{(s)}(t)\}$ in Eq. \eqref{eq:costeffectExpect3}, the estimand for the QALY can thus be defined as 
\begin{eqnarray*}
    E_X \left [ \int_{0}^t E\{ I(T^{(s)}\geq u,D^{(s)}\geq u)|X\}du \right ]  +E_X \left [ \int_{0}^t E\{ I(T^{(s)}\leq u,D^{(s)}\geq u)|X\}Q(u) du \right ].
\end{eqnarray*}
The first term remains the same as the first term in the RMST. However, the second term differs from the RMST as utility loss attributed to the disease is incorporated. 

We can also extend the QALY concept to cause-specific QALY lost due to the disease as $E_X \left \{ \int_{0}^t \int_0^ u f_{ D^{(s)}}( v| T^{(s)}\geq v, X) \Pr( T^{(s)}\geq v|X )dv Q(u) du  \right \}
$, which can be estimated by 
\begin{align*}
    \label{eq:qalossCRC}
   \frac{1}{n}\sum_{i=1}^n \int_0^t\int_0^u \widehat{P}_1(v|X_i,s)\{ 1-\widehat{P}_{13}(u|v,X_i,s) \} d\widehat{\Lambda}_{01}(v|Z(v;S_i=s),X_i)Q(u) du.\vspace{-7mm}
\end{align*}

For an individual having the disease, here, in our CRC example, the quality score $Q(\cdot)$ may vary depending on which phase of cancer care the individual is in. We further illustrate estimation for the QALY and quality adjusted life years lost using utility losses due to cancer care in the real data analysis in Section \ref{Sec:Application}.

\subsubsection{Example 4: Number of Screenings} \label{sec:scr}
The estimand for the number of screenings is a special case of Eq. \eqref{eq:costeffectExpect3} by setting $W_1(u)$ to be the cumulative function for the number of screenings up to time $u$ and $W_2(\cdot)=0$,   because screening, as preventive measure, is assumed to take place before the cancer diagnosis or any other terminal events. In the example of CRC colonoscopy screening, the US Preventive Services Task Force recommends that individuals should commence screening at the age of 50 and undergo colonoscopy every 10 years \citep{bibbins2016screening}. Hence for CRC screening, we can write $W_1(u;s) = \lfloor (u-s+10)_{+}/10 \rfloor$, where $u$ denotes the age in years, $s$ is the age at the first screening (e.g., 50 years old), $x_+$ is $x$ if $x>0$ and 0 otherwise, and $\lfloor x \rfloor$ is the largest integer below $x$. The estimand is  $E\{M^{(s)}(t)\}  =  E_X \int_{0}^t E\{ I(T^{(s)}\geq u,D^{(s)}\geq u)|X\}dW_1(u; s)$, and the cost of screening  can be estimated by 
$ \widehat E\{M^{(s)}(t)\} =  \frac{1}{n}\sum_{i=1}^n \int_{0}^t \widehat{P}_1(u|X_i,s) dW_1(u;s). $ 

\section{Cost-Effectiveness Analysis of CRC Screening} \label{Sec:Application}

We assessed the cost-effectiveness of CRC screening using data from the WHI. Our primary aim is to examine how the starting age for CRC screening affects CRC incidence and mortality. To accomplish this, we conducted the analysis on the 65,062 study participants who had not undergone CRC screening at the beginning of the study. Comprehensive data on socio-demographic and epidemiologic factors were collected 
at study entry. The mean follow-up duration was 16 years with maximum 25 years. During this period, 1,291 individuals ($2.0\%$) developed CRC, of whom 603 ($1.0\%$) died after being diagnosed with CRC. Additionally, 15,485 participants ($23.8\%$) died without experiencing CRC. Out of the total participants, 44,912 ($69.0\%$) underwent CRC screening, with an average age at start screening of 67.1 years.
The outcomes of interest were the age at diagnosis of CRC and age at death, both subject to left truncation (age at enrollment) and right censoring (age at the last follow-up). We employed our proposed estimators to assess RMST, QALY, total cost in terms of the number of colonoscopies, and (quality-adjusted) mean life years lost due to CRC and other causes. These estimators were assessed for different ages at initiating CRC screening, specifically, $s=$ 50, 60, and 70 years, over time interval from 50 years to $t= 60, 70$, and $80$ years. 

Six baseline risk factors were included, based on risk prediction models developed by \cite{freedman2009colorectal} and \cite{liu2014estimating}. These factors were obesity (BMI $<30$, $\geq 30 \text{kg/m}^2$), family history of CRC among first-degree relatives (no, yes), exercise (0, 0-2, $>2$ hours per week), use of aspirin and other nonsteroidal anti-inflammatory drugs (NSAIDs) (nonuser, regular user), vegetable consumption (\# of servings per day), and estrogen status within the last 2 years (negative, positive). Two demographic factors were also included: education (high schoolor less, some college, college/graduate degree) and race and ethnicity (White, Asian/Pacific Islander, Black/African American, Hispanic/Latino, and others). The reference level for each categorical covariate is defined as the first value within the parentheses. SM Table S1 summarizes these risk factors.

We fit the multi-state models (\ref{eq:multimodels1}) and (\ref{eq:multimodels3}) in Section \ref{sec:notation} with time-dependent screening status, while adjusting for the aforementioned demographic and risk factors. In model \eqref{eq:multimodels3} for mortality since cancer, age at diagnosis was also included. The risk sets were adjusted for left truncation by age at enrollment. The results are presented in Table~\ref{tab:tabreg1t}. CRC screening reduced the risk of developing CRC by $57\%$.
Obese Individuals or those with a positive family history had a higher risk of developing CRC, while regular exercise and a positive estrogen status were associated with reduced risk. 
For death after CRC, screening decreased the risk by 17\%. Obesity remained a significant risk factor for mortality, and older age at CRC diagnosis was associated with a higher risk of earlier mortality. 
\begin{table}[!h]
\caption{Summary results of the multi-state models for the association of time-dependent screening and risk factors with disease risk and mortality} \label{tab:tabreg1t}
\renewcommand{\arraystretch}{1.1}
\centering
\scalebox{0.72}{
\begin{tabular}{l|cc|cc|cc}
  \hline
  & \multicolumn{2}{|c|}{Healthy to CRC} &\multicolumn{2}{|c
 |}{Healthy to death }&\multicolumn{2}{|c
 }{CRC to death} \\ \cline{2-3}\cline{4-5} \cline{6-7}
 & HR (95\% CI)$^*$ &p-value& HR (95\% CI) &p-value& HR (95\% CI) &p-value \\
  \hline
 Screening &
0.43 (0.38-0.48) & $<$1.0e-3 & 0.93 (0.90-0.96) & $<$1.0e-3 &  0.83 (0.69-0.99) & 0.04 \\
Obesity& 1.25 (1.11-1.40) & $<$1.0e-3&1.31 (1.26-1.35)&$<$1.0e-3 & 1.22 (1.03-1.44)& 0.02 \\
  Family hx & 1.37 (1.18-1.59) & $<$1.0e-3& 1.04 (0.99-1.08) & 0.13& 0.97 (0.78-1.21) & 0.78 \\
  Exercise & 0.90 (0.84-0.97) & $<$1.0e-3 & 0.93 (0.91-0.95) & $<$1.0e-3 & 0.99 (0.89-1.10) & 0.88\\
  NSAIDS use& 0.93 (0.82-1.05) & 0.23 & 1.14 (1.11-1.18) &$<$1.0e-3 &  1.07 (0.90-1.27) & 0.45 \\
  Hormone use & 0.82 (0.73-0.92) & $<$1.0e-3 & 0.93 (0.90-0.96)& $<$1.0e-3 & 1.03 (0.87-1.23) & 0.72 \\
  Vegetable intake & 1.00 (0.95-1.04) & 0.84 & 0.97 (0.96-0.99) & $<$1.0e-3&0.95 (0.89-1.02) & 0.16 \\
  Racial or ethnic group&&&&\\
 \quad  Asian/Pacific Islander  & 1.26 (0.91-1.74)& 0.17 & 0.78 (0.69-0.89) & $<$1.0e-3&   0.73 (0.41-1.30)& 0.28 \\
 \quad  Black/African American& 1.15 (0.95-1.39) & 0.16 & 1.12 (1.05-1.19) &$<$1.0e-3 &1.03 (0.74-1.42) & 0.87 \\
   \quad  Hispanic/Latino& 0.87 (0.63-1.19) & 0.38 & 0.85 (0.77-0.95) & $<$1.0e-3 & 0.95 (0.58-1.56) & 0.84\\
  \quad  Others & 1.02 (0.66-1.59) & 0.93 & 1.08 (0.95-1.23) & 0.24 & 0.83 (0.41-1.67) & 0.59\\
Education&&&&\\
  \quad Some college & 1.04 (0.91-1.20) & 0.55 &0.99 (0.95-1.03) & 0.62&0.99 (0.81-1.22) & 0.95\\
  \quad College/graduate degree & 1.00 (0.87-1.16)& 0.98 & 0.89 (0.85-0.93)& $<$1.0e-3& 0.83 (0.67-1.03) & 0.10 \\
  Age at diagnosis (year) &&&&& 1.07 (1.06-1.08) & $<$1.0e-3
  \\
  \hline
  \multicolumn{7}{l}{\footnotesize{$^*$ HR (95\%CI): hazard ratio (95\% confidence interval)}}
\end{tabular}
}
\end{table}

\subsection{Restricted Mean Survival Time (RMST)}
Based on these models we estimated the RMST for no screening and screening at age 50, 60 and 70 years old (Table \ref{tab:rmstWHI2}(a)). We used 100 bootstrap samples to calculate the standard errors (SE). Throughout this section, we report life years gained per 1000 individuals, following the convention of cost-effectiveness analyses (see, e.g., \citealt{meester2018optimizing}). Compared to no screening, initiating screening at age 50 would improve the RMST over a 30-years follow-up period ($t=80$) for the disease-free stage by (28.408 - 28.098)*1000  = 310 years per 1000 individuals (p-value $< 0.001$, 95\% CI, 196 to 424). On the other hand, it would reduce RMST (0.142 - 0.309)*1000 = -167 years (95\% CI, -208 to -126, p-value $<0.001$) for individuals who developed CRC, as those who underwent screening and still developed CRC tended to have cancer at an older age, resulting in shorter RMST since cancer diagnosis. Importantly, the total RMST would still  improve by (28.550-28.407) * 1000 = 143 years per 1000 individuals (95\% CI,  39 to 247, p-value $=0.005$). 
Screening at older ages, specifically at 60 and 70 years, would also lead to improvements in both total RMST and disease-free RMST, albeit to a lesser degree. Similar benefits were observed with shorter follow-up times at $t=60$ and $70$ years.


\begin{table}[htbp]
\caption{Summary of the RMST for $t= 60, 70$, and $80$ years for no screening (No Scr) and screening at age 50 (Scr 50), 60 (Scr 60) and 70 (Scr 70) years.} \label{tab:rmstWHI2}
\renewcommand{\arraystretch}{1.1}
\centering
\scalebox{.77}{
\begin{tabular}{l cccc cccc cccc}
  \hline
 \multicolumn{13}{l}{(a) Proposed Multi-State Approach} \\ \hline
  &  \multicolumn{4}{c}{Disease-free} & \multicolumn{4}{c}{Disease to death} &
  \multicolumn{4}{c}{Total RMST} \\
  RMST & \multicolumn{1}{c}{No Scr} & Scr 50& Scr 60 & Scr 70& No Scr& Scr 50& Scr 60 & Scr 70&No Scr& Scr 50& Scr 60 & Scr 70 \\
  \hline
 \multicolumn{13}{l}{\it t = 60}\\
Est.&   9.916 & 9.934 & -* & - &  0.021 & 0.009 & - & - & 9.937 & 9.942 &  - & - \\
  SE & 0.011 & 0.008 & - & - & 0.005 & 0.003& - & - & 0.008 & 0.008 &- & - \\
 
\hline
\multicolumn{13}{l}{\it t = 70}\\
  Est.& 19.449 & 19.551 & 19.495 & -$^{**}$ & 0.118 & 0.053 & 0.088& - & 19.567 & 19.604 & 19.584 & - \\
  SE & 0.025 & 0.022 & 0.026 &-& 0.012 & 0.006 & 0.012&- & 0.021 & 0.021 & 0.021 &- \\
  \hline
\multicolumn{13}{l}{\it t = 80}\\

  Est.&  28.098 & 28.408 & 28.317 & 28.175 & 0.309 & 0.142 & 0.194 & 0.271 & 28.407 & 28.550 & 28.511 & 28.446 \\
  SE & 0.042 & 0.040 & 0.043 & 0.042 & 0.018 & 0.011 & 0.017 & 0.018 & 0.037 & 0.038 & 0.038 & 0.036 \\
  \hline 
 \multicolumn{13}{l}{(b) Overall Mortality Approach} \\ \hline
  RMST & No Scr& Scr 50& Scr 60 & Scr 70 & & & & & & & & \\
  \hline
 \multicolumn{13}{l}{\it t = 60}\\
Est.& 9.936 & 9.941  & -- & -- & & & & & & & &\\
  SE & 0.009 & 0.008& -- & -- & & & & & & & &\\
\hline
\multicolumn{13}{l}{\it t = 70}\\
  Est. & 19.560 & 19.598 & 19.578& -- & & & & & & & &\\
  SE & 0.021 & 0.021 & 0.022 & -- & & & & & & & &\\
   \hline
\multicolumn{13}{l}{\it t = 80}\\
  Est& 28.409 & 28.544 & 28.510 & 28.452& & & & & & &\\
  SE   & 0.036 & 0.038 & 0.038& 0.036  & & & & & & & & \\
  \hline
  \multicolumn{13}{l}{\footnotesize
$^*$ Screening at 60 or 70 for $t = 60$ is same as no screening. \quad 
$^{**}$ Screening at 70 for $t=70$ is same as no screening. 
}
\end{tabular}}
\vspace{-4mm}
\end{table}

For comparison, we fit a Cox proportional hazards model treating death as a single primary event. We used the same six risk factors as in the multi-state models (SM Table~S2), and the associations were consistent with those observed for mortality before CRC in the multi-state modeling approach. The RMST estimates under this model were in line with the total RMST obtained from the multi-state model (Table~\ref{tab:rmstWHI2}(b)). For example, the RMST for screening at 50 years at $t=80$ was 28.544 vs. the total RMST 28.550 under the multi-state model. This similarity is likely due to the fact that CRC is relatively rare, and the majority would die without CRC. When the disease is not rare, our multi-state modeling-based estimators are more efficient and less biased, as illustrated in one of the simulation scenarios in Section~\ref{Sec:Simulation}.

\subsection{Quality-adjusted Life Years (QALY)} \label{sec:qaly}

For evaluating the cost-effectiveness of cancer screening, studies often report QALY to account for the generally significantly lower health quality during the cancer stage \citep{ratushnyak2019cost}. Our proposed measure can incorporate the quality of life score as a time-varying weight function to obtain QALY. Specifically, for the cancer-free stage, we set $Q(u) = 1$, the same as the RMST for this stage, as there was no loss of quality of life during the disease-free stage. For the cancer stage, we derived the quality of life score based on Table 1 in \cite{van2014should}. Briefly, we considered three clinically relevant phases for CRC care: initial, continuing, and terminal care. The initial care phase was defined as the first year after diagnosis; the terminal care phase was the last year of life; and the continuing care phase was all years in between. For participants surviving $\leq 2$ years, the final year was considered the terminal care, and the remaining time was  allocated to initial care. For participants surviving $\leq 1$ year, all of the time was considered the terminal care. The quality score for CRC care is measured on a continuous scale and depends on the duration between the age at cancer diagnosis and age at death. Let $a$, $b$ and $c$ be the quality scores associated with the initial, continuing and terminal care for CRC, respectively (SM Table S3), and we have:  
\begin{eqnarray*}
Q(u; T, D) & = & \left \{
\begin{array}{ll}
aI(T\leq u \leq T+1)+bI(T+1<u\leq D-1) & \\
\mbox{} \qquad + cI(D-1<u\leq D), & D-T > 2,\\
aI(T\leq u \leq D-1)+cI(D-1<u\leq D), & 1 < D-T \geq 2,\\
cI(T<u\leq D), & D-T \leq1.
\end{array}
\right .
\end{eqnarray*}
We incorporated $Q(u; T, D)$ and obtained $\widehat{E}\{ M(t) \}$ accordingly (details are provided in SM Section 4.1), and the results are presented in Table \ref{tab:QALYWHI2}. The difference in QALY during the cancer-free stage was the same as that of RMST. However, the difference in QALY during the cancer stage between individuals screened at age 50 years old and no screening would be 121 years  per 1000 individuals (95\% CI, 92 to 150). The overall QALY gain from screening at age 50 compared to no screening would be 189 years per 1000 individuals (95\% CI, 83 to 295) , which was greater than the increase simply in life years ($\sim \! 143$ years per 1000 individuals), as indicated by the RMST. 
\begin{table}[htbp]
\caption{Summary of the QALY at $t=60$, $70$, and $80$ years for no screening (No Scr) and for screening at age 50 (Scr 50), 60 (Scr 60) and 70 (Scr 70) years.} \label{tab:QALYWHI2}
\renewcommand{\arraystretch}{1.1}
\centering
\scalebox{.77}{
\begin{tabular}{l|cccc|cccc|cccc}
  \hline
 QALY &  \multicolumn{4}{c|}{Disease-free} & \multicolumn{4}{c|}{Disease to death} &
  \multicolumn{4}{c}{Total QALY} \\
 & \multicolumn{1}{c}{No Scr} & Scr 50& Scr 60 & Scr 70& No Scr& Scr 50& Scr 60 & Scr 70&No Scr& Scr 50& Scr 60 & Scr 70 \\
  \hline
 \multicolumn{13}{l}{\it $t = 60$}\\
 \hline
Est.&   9.916 & 9.934 & -* & - & 0.015 & 0.007 & -&-& 9.931 & 9.941 & - & -\\
  SE & 0.011 & 0.008 & - & -  & 0.004 & 0.002 & - & - & 0.008 & 0.007 &- & - \\
\hline
\multicolumn{13}{l}{\it $t = 70$}\\
 \hline
  Est.&19.449 & 19.551 & 19.495 & -$^{**}$  & 0.085 & 0.038 & 0.064 & - & 19.535 & 19.589 & 19.559 & -\\
  SE & 0.025 & 0.022 & 0.026& - & 0.009 & 0.004 & 0.008& -& 0.021 & 0.021 & 0.022& -\\
   \hline
\multicolumn{13}{l}{\it $t = 80$}\\
\hline
  Est. & 28.098 & 28.408 & 28.317 & 28.175 & 0.224 & 0.103 & 0.140 & 0.196 & 28.322 & 28.511 & 28.457 & 28.371 \\
  SE. & 0.042 & 0.040 & 0.043 & 0.042 & 0.013 & 0.008 & 0.012 & 0.013 & 0.037 & 0.039 & 0.039 & 0.037 \\
  \hline
  \multicolumn{13}{l}{\footnotesize
$^*$ Screening at 60 or 70 for $t = 60$ is same as no screening. \quad 
$^{**}$ Screening at 70 for $t=70$ is same as no screening. 
}
\end{tabular}}
\end{table}

\subsection{Cause-Specific (Quality-Adjusted) Mean Life Years Lost}

An important advantage of multi-state modeling is its ability to quantify mean life years lost due to different causes, such as CRC and other causes. This helps us understand the impact of CRC screening on preventing CRC and, consequently, reducing the years lost due to CRC. This is particularly valuable when the disease incidence rate is low. If we only examine the total RMST or QALY, the impact of screening on disease prevention may be overlooked. As shown in Table \ref{tab:lostWHI2} and SM Figure S1, if the population underwent screening at age 50, there would be a significant reduction in years lost due to CRC (0.099-0.037)/0.099 = 62.6\% (p-value$<0.001$) during the 30-years follow-up, compared to no screening. Screening at older age would also reduce mean years lost to CRC, but to a lesser extent. There were no statistically significant differences in years lost due to other causes between those who had screening  and those who did not (p-value $> 0.05$). When accounting for the quality score, screening  would similarly reduce QALY lost due to CRC compared to no screening (Table~\ref{tab:lostWHI2} and SM Figure S2). 

 
\begin{table}[htbp]
\caption{ Summary of (quality-adjusted) cause-specific mean life years lost at $t=60$, $70$, and $80$ years for no screening (No Scr) and screening starting at ages 50 (Scr 50), 60 (Scr 60), and 70 (Scr 70) years.} \label{tab:lostWHI2}
\renewcommand{\arraystretch}{1.1}
\centering
\scalebox{0.79}{
\begin{tabular}{l|cccc|cccc|cccc}
  \hline
&\multicolumn{4}{c|}{Other Causes}&\multicolumn{4}{c|}{CRC} &\multicolumn{4}{c}{Quality-Adjusted CRC}  \\
Years lost &{No Scr}&{Scr 50}&{Scr 60}&{Scr 70}&{No Scr}&{Scr 50}&{Scr 60}&{Scr 70} &{No Scr}&{Scr 50}&{Scr 60}&{Scr 70} \\
  \hline
\multicolumn{13}{l}{\it $t = 60$}\\
 \hline
 Est& 0.061 & 0.057 & --* & -- & 0.002 & 0.001 & -- & --  & 0.001 & 4e-4 &-- & --\\
  SE & 0.008 & 0.008 &  -- &  -- & 0.001 & 2e-4 &  -- & --  & 4e-4 & 2e-4 & -- & -- \\
   \hline
\multicolumn{13}{l}{\it $t = 70$}\\
 \hline
  Est &  0.418 & 0.392 & 0.406 &  --** & 0.018 & 0.006 & 0.013 &  -- &0.014 & 0.005 & 0.010 & --\\
  SE . & 0.020 & 0.021 & 0.021 &  --& 0.003 & 0.001 & 0.003 &  -- & 0.002 & 0.001 & 0.002 & -- \\
   \hline
\multicolumn{13}{l}{\it $t = 80$}\\
 \hline
  Est  &  1.509 & 1.426 & 1.446 & 1.480 & 0.099 & 0.037 & 0.056 & 0.086&  0.076 & 0.028 & 0.043 & 0.066 \\
  SE & 0.034 & 0.038 & 0.037 & 0.034 & 0.011 & 0.004 & 0.008 & 0.010& 0.008 & 0.003 & 0.006 & 0.008\\
    \hline
   \multicolumn{13}{l}{\footnotesize
$^*$ Screening at 60 or 70 for $t = 60$ is same as no screening. \quad 
$^{**}$ Screening at 70 for $t=70$ is same as no screening. 
}
\end{tabular}
}
\end{table}

\subsection{Cost-Effectiveness Analysis}

\begin{figure}
    \centering
\includegraphics[scale=0.35]{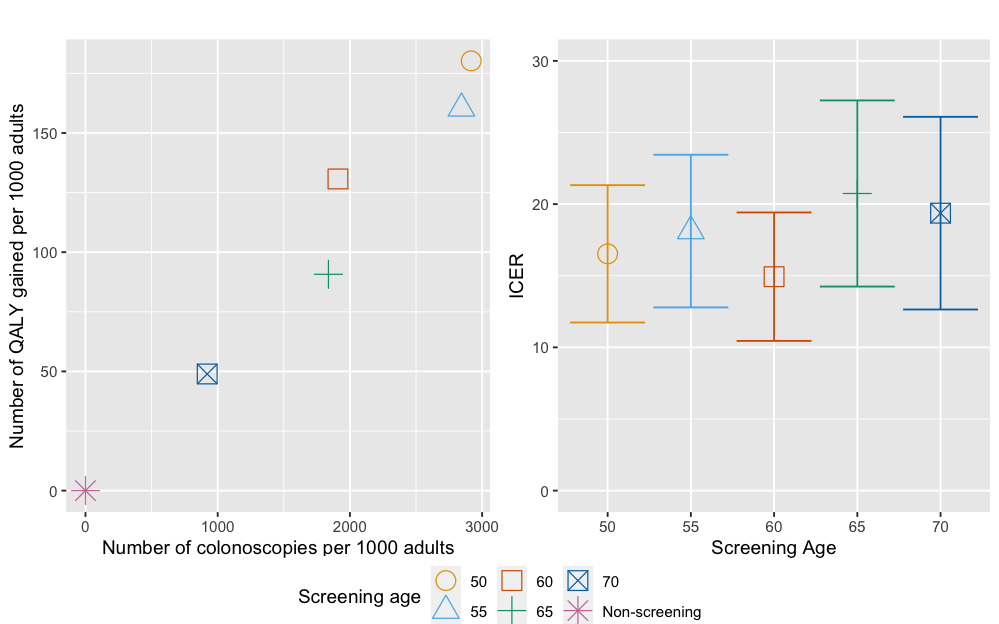}
    \caption{Scatterplot of QALY gained vs. total number of screenings per 1000 individuals (left panel) and ICER for initial screening ages from 50 to 70 years (right panel).}
    \label{fig:qalycost}
\end{figure}

Although earlier screening extends lifetime and reduce mean life years lost due to CRC, itimposes a burden on individuals and strains societal resources. This is because earlier screening results in more screenings over a lifetime, following the guidelines of a colonoscopy every 10 years for CRC screening \citep{bibbins2016screening}. To understand the (quality-adjusted) life years gained relative to the number of screenings, we estimated the number of screenings assuming individuals would adhere to CRC guidelines as described in Section~\ref{sec:scr}. As expected, the earlier the initial screening age, the greater the total number of screenings during the follow-up (SM Table S4) and the greater QALY gain (left panel in Figure \ref{fig:qalycost}). We calculated the incremental cost-effectiveness ratio (ICER), dividing the increments in the number of screenings by the increments in QALY compared to no screening (right panel in Figure \ref{fig:qalycost}). When the WHI was initiated, CRC screening was not widely in the US, and it was more than a decade later when the US Preventive Services Task Force first recommended starting screening at age 50. In this context, we considered no screening as the benchmark strategy. We considered initial screening ages at 50, 55, 60, 65 and 70 years, adding 55 and 65 years to provide a clearer trend analysis. The incremental cost-effectiveness ratio ranged from 13.9 to 18.9 on average. Following the principle that recommendable strategies should have an acceptable ICER (typically $<$50) \citep{meester2018optimizing}, all strategies have an acceptable ICER, with screening at age 50 resulting in the highest QALY. This finding provides the real-world evidence in support of colonoscopy screening for CRC at age 50 years.


We assessed the proportional hazards assumptions of all models, and model (\ref{eq:multimodels3}) shows a modest violation (p-value = 0.04). A close examination suggests the violation is driven by the non-proportional effect of age at diagnosis (p-value $<0.01$, SM Figure S3). We thus fit a natural cubic spline to reflect the time-varying effect of age at diagnosis (SM Figure S5 and SM Section 4.4). There was little impact on the regression coefficient estimates of all other risk factors (SM Table S5) and the RMST, QALY and cause-specific mean life years lost estimates for screening at various ages and no screening were similar, though with larger standard errors than the multi-state proportional hazards models (SM Table S6). 


\section{Simulation} \label{Sec:Simulation}
We conducted simulation studies to examine the finite-sample performance of the proposed multi-state modeling approach for measuring cost-effectiveness. We considered two simulation settings. The first setting (Setting I) aims to demonstrate the finite-sample properties of our proposed approach by modelling the entire disease progression process under a wide range of scenarios. The second setting (Setting II) is designed to mimic the real-data example, in which the disease incidence is relatively low. In each simulation setting, we evaluated the RMST and cause-specific mean years lost due to the disease or other causes for no screening and screening at age 50 years.  

\subsection{Simulation Setting I}
In this setting, we generated data under a wide range of scenarios. 
Specifically, we used the Weibull model for the transition hazard functions  $\lambda_{01}(u|Z(u;S),X)={5.2}/{56.3^{5.2}} t^{5.2-1}\exp\{-1.4Z(u;S)+0.5X\}$ and $\lambda_{02}(u|Z(u;S),X)=$ 
${5.9}/{83.0^{5.9}} t^{5.9-1}\exp\{-0.05Z(u;S)+0.4X\}$. Here, $X$ was a risk score that may include lifestyle and environmental risk factors (e.g., smoking, alcohol) and genetic risk predispostion (e.g., polygenic risk score). While individual factors in a risk score may be discrete, the aggregation of these risk factors  was typically continuous. For simplicity, we generated $X \sim \textup{Normal}(0,1)$. The screening age $S$ was generated as a function of an individual's risk score $X$ from an exponential distribution with mean $50\exp(0.53X)$. For those who developed the disease,  the duration between disease onset and death was generated from an exponential distribution with mean $45\exp(-0.3X+0.05T)$. For the censoring distribution, we considered two scenarios: (i) the censoring time $C$ was completely independent of $T$, $D$, $S$ and $X$, following a $\textup{Uniform}(40,100)$ distribution; (ii) $C$ was conditionally independent of $T$ and $D$ given $X$, and it  was generated under the Cox model with the baseline hazard function assumed to follow the US population age distribution and hazard ratio $\exp(X)$. Censoring scenario (i) resulted in approximately 25.4\% of individuals developing the disease, 5.0\% experiencing death before disease onset, and an overall death rate of $21.2\%$. Censoring scenario (ii) yielded similar incidence and death rates. The sample size was set to be $n=2,500$ and a total of 300 datasets were generated under each scenario. To evaluate the performance of bootstrap-based inference, 50 bootstrap samples were generated for each
simulated dataset.

\begin{table}[!h]
\caption{Summary of RMST estimates over the age range $[40,70]$ for screening at 50 years (Scr) and no-screening (No Scr). } 
\label{tab:simRMST}
\renewcommand{\arraystretch}{1.1}
\centering
\scalebox{0.80}{
\begin{tabular}{l cccc cccc cccc}
  \hline
  \multicolumn{13}{l}{\it Scenario (i): Independent censoring}\\
  \cline{1-6}
  \multicolumn{13}{l}{(a) Proposed multi-state approach} \\ \hline
    & \multicolumn{4}{c}{Disease-free}
 & \multicolumn{4}{c}{Disease to death}
 &\multicolumn{4}{c}{Total}\\
 \cline{3-4}\cline{7-8}\cline{11-12}
  &\multicolumn{2}{c}{True}&\multicolumn{2}{c}{Proposed} &\multicolumn{2}{c}{True}&\multicolumn{2}{c}{Proposed} &\multicolumn{2}{c}{True}&\multicolumn{2}{c}{Proposed}\\
  & No Scr &Scr &No Scr&Scr
  & No Scr&Scr&No Scr&Scr
  & No Scr&Scr
  & No Scr&Scr\\
  \hline
Mean$^*$  &11.833 & 14.646 & 11.812 & 14.617& 3.014 & 2.333 & 3.082 & 2.416 & 14.846 & 16.979 & 14.894 & 17.033\\
  ESD$^*$ &&& 0.470 & 0.454 &&& 0.202 & 0.179 &&& 0.433 & 0.407 \\
  SE$^*$ && &0.458 & 0.448 &&& 0.203 & 0.181 &&& 0.433 & 0.407\\
  CP$^*$ &&& 0.933 & 0.950&& & 0.953 & 0.943 &&&0.937&0.945\\
  \hline
    \multicolumn{13}{l}{(b) Overall mortality approach} \\ \hline
   
   &\multicolumn{2}{c}{True}&\multicolumn{2}{c}{Overall death}\\
  &&&No Scr&Scr\\
  \hline
Mean & 14.846 & 16.979& 14.940 & 17.995&\\
 ESD &&& 0.472 & 0.408&\\
 SE &&& 0.491 & 0.403&\\
 CP &&&0.947&0.300&\\
 \hline
  \multicolumn{13}{l}{\it Scenario (ii): Conditional independent censoring}\\
  \cline{1-6}
  \multicolumn{13}{l}{(a) Proposed multi-state approach} \\ \hline
    & \multicolumn{4}{c}{Disease-free}
 & \multicolumn{4}{c}{Disease to Death}
 &\multicolumn{4}{c}{Total}\\
 \cline{3-4}\cline{7-8}\cline{11-12}
  &\multicolumn{2}{c}{True}&\multicolumn{2}{c}{Proposed} &\multicolumn{2}{c}{True}&\multicolumn{2}{c}{Proposed} &\multicolumn{2}{c}{True}&\multicolumn{2}{c}{Proposed}\\
   & No Scr &Scr &No Scr&Scr
  & No Scr&Scr&No Scr&Scr
  & No Scr&Scr
  & No Scr&Scr\\
  \hline
Mean  & 11.833 & 14.646  &11.813 & 14.611 & 3.014 & 2.333&  3.087 & 2.422& 14.846 & 16.979 & 14.900 & 17.033 \\
 ESD &&& 0.479 & 0.465 &&&0.210 & 0.186 && & 0.436 & 0.405\\
 SE &&&  0.463 & 0.449  &&& 0.208 & 0.184 && & 0.441 & 0.402 \\
  CP&& & 0.947 & 0.937 &&& 0.950 & 0.950 && & 0.950 & 0.943\\
  \hline
   \multicolumn{13}{l}{(b) Overall mortality approach} \\ \hline
   
   &\multicolumn{2}{c}{True}&\multicolumn{2}{c}{overall death}\\

  &&&No Scr&Scr\\
  \hline
Mean& 14.846 & 16.979 & 14.830 & 17.946\\
ESD  && &  0.458 & 0.400 \\
SE  &&& 0.489 & 0.401\\
CP && & 0.947 & 0.333 \\
\hline
\end{tabular}
}
\footnotesize{
\begin{itemize}
    \item[*] Mean, ESD, SE and CP are the mean, empirical standard deviation
of estimates, mean of the bootstrap-based SE, and coverage probability of $95\%$ CI over 300 datasets, respectively.
\end{itemize}
}
\end{table}

 We compared the proposed multi-state approach to an approach that only models the terminal event, death, which we referred to as the 'overall mortality' approach. We evaluated the RMST over the age range $[40,70]$ for both scenarios, initiating screening at 50 years and having no screening. Table \ref{tab:simRMST} presents a summary of the proposed multi-state and overall mortality RMST estimators. Under both censoring scenarios, the proposed estimator exhibited no bias. The means of the bootstrap-based SE estimates closely matched with the empirical standard deviations (ESD), and the coverage probabilities of $95\%$ CI were close to 95\% for both states (disease-free and disease) and overall.  In comparison, the RMST estimates obtained from the overall mortality approach were biased for the screening group with coverage probabilities well below 95\%, with 30.0\% and 33.3\% under the independent censoring and conditional independent censoring scenarios, respectively. For the no-screening group, both the proposed and overall-mortality approaches were unbiased; however, the proposed estimators were more efficient. We also examined the performance of the multi-state approach for estimating the cause-specific mean life years lost due to disease and other causes (Table \ref{tab:simLost2}). Similarly, the proposed estimators for the mean life years lost exhibited no bias. The bootstrap-based SEs closely approximated the ESD and the coverage probabilities were $\sim$ 95\%. 
\begin{table}[htbp]
\caption{Summary of mean life years lost estimates over the age range $[40,70]$ for screening at 50 years (Scr) and no screening (No Scr).} 
\label{tab:simLost2}
\renewcommand{\arraystretch}{1.1}
\centering
\scalebox{0.8}{
\begin{tabular}{l cccc c cccc}
\hline
&\multicolumn{4}{c}{\it Death due to CRC}& &\multicolumn{4}{c}{\it Death due to other causes}\\
\cline{2-5} \cline{7-10}
&\multicolumn{2}{c}{True}&\multicolumn{2}{c}{Proposed}& & \multicolumn{2}{c}{True}&\multicolumn{2}{c}{Proposed}\\
&No Scr&Scr&No Scr&Scr& & No Scr&Scr&No Scr&Scr\\
\hline
\multicolumn{10}{l}{\it Scenario (i): Independent censoring}\\
\hline
Mean & 13.559 & 11.125 & 13.500 & 11.081 && 1.754 & 2.033 & 1.742 & 2.021  \\
ESD  & & & 0.442 & 0.426 &&& & 0.267 & 0.193\\
  SE  & & & 0.454 & 0.429& & & & 0.274 & 0.196\\
  CP & && 0.950 & 0.943 & & & & 0.950 & 0.940 \\
\hline
\multicolumn{10}{l}{\it Scenario (ii): Conditional independent censoring}\\
\hline
Mean & 13.559 & 11.125 & 13.493 & 11.082&& 1.754 & 2.033 & 1.744 & 2.020\\
  ESD & & & 0.445 &0.423 &&  & &   0.284 & 0.199 \\
  SE   &  &  & 0.456 & 0.424& & &  & 0.291 & 0.209 \\
  CP  &  &  & 0.957 & 0.940&&  & & 0.953& 0.950  \\
\hline
\end{tabular}
}
\end{table}
\vspace{-.5cm}

\subsection{Simulation Setting II}
In this setting, we generated the data to mimic the colorectal cancer study described in Section~\ref{Sec:Application}. Specifically, we generated the time to CRC based on a Weibull model fitted to the CRC age-specific
SEER incidence rates and  age at death based on a Weibull model fitted to the US mortality data. All other settings were the same as in Setting I. The simulation results were based on 100 simulated datasets, each with $n=20,000$ individuals, under the conditional independence censoring scenario. On average, approximately $2.0\%$ of individuals developed the disease, $15.0\%$ died before developing the disease, and $1.4\%$ died after developing the disease.

Table \ref{tab:simRMST2} presents a summary of the RMST estimators for both the proposed multi-state and overall mortality approaches. Both the proposed and overall mortality estimators performed well in estimating the total RMST, and for the proposed approaches, the RMST estimators for both states also performed well.  
Both estimators showed that screening extended the total lifetime, although the increments were modest. By modelling the entire process, it can be seen that screening primarily extended the state being disease-free, as it was intended to prevent or delay disease onset. This observation is more evident when examining the mean life years lost due to disease and other causes using our proposed approach (Table \ref{tab:simLost3}). Screening significantly reduced the mean life years lost due to CRC, whereas there was little difference in the mean life years lost due to other causes. These findings were consistent with the analysis results of the WHI colorectal cancer data, where disease incidence was low, and most deaths occurred before the disease occurs. 
\begin{table}[!h]
\caption{Summary of RMST over the age range $[40,70]$ for screening at 50 years (Scr) and for no screening (No Scr) under the conditional independent censoring scenario.} 
\label{tab:simRMST2}
\renewcommand{\arraystretch}{1.1}
\scalebox{0.8}{
\centering
\begin{tabular}{l llll llll llll}
  \hline
  \multicolumn{13}{l}{(a) Proposed multi-state approach} \\ \hline
    & \multicolumn{4}{c}{Disease-free}
 & \multicolumn{4}{c}{Disease to death}
 &\multicolumn{4}{c}{Total}\\
 \cline{3-4}\cline{7-8}\cline{11-12}
  &\multicolumn{2}{c}{True}&\multicolumn{2}{c}{Proposed} &\multicolumn{2}{c}{True}&\multicolumn{2}{c}{Proposed} &\multicolumn{2}{c}{True}&\multicolumn{2}{c}{Proposed}\\
  & No Scr &Scr &No Scr&Scr
  & No Scr&Scr&No Scr&Scr
  & No Scr&Scr
  & No Scr&Scr\\
  \hline
  
Mean & 21.483 & 22.006& 21.487 & 22.016 & 0.786 & 0.514& 0.782 & 0.516 & 22.269 & 22.520 & 22.269 & 22.532 \\
ESD &&& 0.188 & 0.123 &&& 0.063 & 0.041 && & 0.177 & 0.115 \\
SE&&& 0.198 & 0.123 &&& 0.066 & 0.040 && & 0.189 & 0.120 \\
 CP&&  & 0.950 & 0.960 &&& 0.960 & 0.930 && & 0.950 & 0.980 \\
   \hline
    \multicolumn{13}{l}{(b) Overall mortality approach} \\ \hline
   
   &\multicolumn{2}{c}{True}&\multicolumn{2}{c}{Overall death}\\

  &&&No Scr&Scr\\
  \hline
  Mean & 22.269 & 22.520 & 22.446 & 22.731\\
ESD &&&  0.179 & 0.118\\
SE&&& 0.191 & 0.122\\
CP&&  & 0.950 & 0.960\\
\hline
\end{tabular}}
\footnotesize{
}
\end{table}

\begin{table}[!h]
\caption{Summary of mean life years lost over $[40,70]$ for screening at 50 years (Scr) and no screening (No Scr).} 
\label{tab:simLost3}
\renewcommand{\arraystretch}{1.2}
\centering
\scalebox{0.8}{
\begin{tabular}{l ll ll c ll ll}
\hline
&\multicolumn{4}{c}{\it Death due to disease} & &\multicolumn{4}{c}{\it Death due to other causes}\\
\cline{2-5} \cline{7-10}
&\multicolumn{2}{c}{True}&\multicolumn{2}{c}{Proposed}& & \multicolumn{2}{c}{True}&\multicolumn{2}{c}{Proposed}\\
&No Scr&Scr&No Scr&Scr& & No Scr&Scr&No Scr&Scr\\
\hline
Mean  & 0.721 & 0.518 & 0.714 & 0.510 & & 7.007 & 6.958 & 7.019 & 6.953 \\
  ESD& && 0.059 & 0.041&  & &  & 0.183 & 0.113  \\
  SE  && & 0.062 & 0.041&& & & 0.190 & 0.117 \\
  CP  & & & 0.970 & 0.950& & & & 0.950 & 0.960 \\
\hline
\end{tabular}
}
\vspace{-5mm}
\end{table}

\section{Conclusions}
In this paper, we develop a causal framework for evaluating the cost-effectiveness of a possibly censored continuous intervention. Our approach incorporates a unified measure that encompasses common measures such as restricted mean survival time and quality adjusted life years. The unified measure also includes cost-related measurements, such as the number of screenings during the follow up. We employ a multi-state model and treat the data as semi-competing risks. In contrast to conventional cost-effectiveness analysis approaches that focus solely on modeling the terminal event, our multi-state modeling approach allows for the estimation for benefit and costs at each state during disease progression, which can better capture and disentangle the effect of an intervention on the target diseases.  We establish the asymptotic properties, namely, consistency and asymptotic normality, of proposed estimators. An application to a  colorectal cancer study shows that cancer screening improves overall life expectancy, particularly healthy life years. Notably, our approach indicates that initiating screening at age 50 would offer the greatest improvement compared to no screening within the acceptable ICER. Extensive simulation results  show that our proposed multi-state estimators are more accurate and efficient compared to conventional approaches that primarily focus on the terminal event. 

The Cox proportional hazards model is a well-studied and widely used model in biomedical applications due to its practical advantages and established utility. However, the Cox model's validity hinges on the proportional hazards assumption, which, if violated, can compromise the accuracy of the results. In our application, we meticulously evaluated the goodness-of-fit for models \eqref{eq:multimodels1} and \eqref{eq:multimodels3}. We identified a potential violation of the proportional hazards assumption in the sojourn model \eqref{eq:multimodels3}, specifically concerning the effect of age at diagnosis. To address this, we incorporated a time-varying effect using splines for age at disease. Despite this adjustments, the results remained largely unchanged. To accommodate more complex relationships between confounders, treatment effects, and survival outcomes, one may consider extending the analysis using survival trees \citep{ishwaran2008random}. Alternatively, one may consider incorporating the effect of confounders through the inverse probability weighting \citep{yang2018modeling}. This approach, however, necessitates modeling the time-varying intervention as a function of confounders, a process complicated by semi-competing risks data. Future research in this area could provide valuable insights.  

It is important to clarify that this paper focuses on the cost-effectiveness of cancer screening for preventive purposes, as opposed to screening for early detection. Screening for early detection involves detecting cancer that has already developed but is not yet clinically apparent, a stage known as the pre-clinical phase. Early detection generally leads to better prognoses and more effective treatment options. However, this could introduce length bias, as individuals with longer pre-clinical phase may have a prolonged window for detection, potentially leading to over-representation in screening outcomes. Additionally, some individuals in the pre-clinical phase may never progress to advanced stages, resulting in unnecessary often-intensive cancer treatment. In contrast, preventive screening aims to prevent cancer from occurring in the first place, for instance, by removing precursor lesions before they can develop into cancer. For example, in colorectal cancer, lesions are removed during colonsocopy screenings, negating the need for further treatment. Preventive screening targets healthy individuals with no current cancer, making length bias irrelevant in this context.  

The proposed multi-state model may be overly simplistic for describing the carcinogensis process. Microsimulation modeling, which simulates the disease process for an individual based on a natural history model, can more accurately reflect the adenoma-carcinoma sequence that describes the progression from benign adenomatous polyps to malignant colorectal cancer. However, some underlying assumptions of this approach cannot be directly verified. Incorporating the adenoma-carcinoma sequence into multi-state model is challenging, because precursor lesions such as polyps or adenomas detected during colonoscopy, are removed, thereby truncating their growth as intended. Our method could be expanded to include surveillance screening with follow-up screenings. However, due to the complexity (such as truncated growth of precursor lesions and dependent screening schedules), this would require separate development, which is beyond the scope of this paper.

Cost-effectiveness analysis for cancer preventive screening is complex and requires different considerations than those for early detection or treatment interventions. Microsimulation modeling has been the primary tool in this area. However, observational data, such as those from the WHI, despite their complexity and limitations, offer real-world evidence and provide invaluable insights for assessing the cost-effectiveness of cancer preventive screening. Empirical modeling should be viewed as complementary to microsimulation modeling. In this context, our proposed causal inference framework for cost-effectiveness analysis of time-varying screening (here, age at first screening) is a starting point to address these important questions using real data.  


\section*{ACKNOWLEDGEMENTS}
	
The work is supported in part by the grants from the National Institutes of Health (R01 CA189532, R01 CA195789, R01 CA236558, P30 CA015704, and U01 CA86368) and the Scientific Computing Infrastructure at the Fred Hutchinson Cancer Research Center which is funded by ORIP grant S10OD028685.  
The authors are grateful to the generosity of WHI investigators for using the WHI data to illustrate the proposed method.
The WHI program is funded by the National Heart, Lung, and Blood Institute, National Institutes of Health, U.S. Department of Health and Human Services through contracts HHSN268201600018C, HHSN268201600001C, HHSN268201600002C, HHSN268201600003C, and HHSN268201600004C. 

\bibliographystyle{apalike}

\bibliography{cost}
\end{document}